\documentclass[letterpaper, 10 pt, conference]{ieeeconf}  

\IEEEoverridecommandlockouts                              
\usepackage{acronym}
\newacro{ODE}{ordinary differential equation}
\newacro{PINN}{physics-informed neural network}
\newacro{KBINN}{Kalman-Bucy-informed neural network}
\newacro{SPINN}{stochastic physics-informed neural network}
\newacro{BNN}{Bayesian neural network}
\newacro{KBF}{Kalman-Bucy filter}
\newacro{EKBF}{extended Kalman-Bucy filter}

\usepackage{csquotes}
\usepackage{balance}

\title{\LARGE \bf
	Kalman-Bucy-Informed Neural Network for System Identification
}

\author{Tobias Nagel and Marco F. Huber
	\thanks{Tobias Nagel and Marco F. Huber are with the Fraunhofer Institute for Manufacturing Engineering and Automation IPA, Center for Cyber Cognitive Intelligence (CCI), 70569 Stuttgart, Germany {\tt\{tobias.nagel, marco.huber\}@ipa.fraunhofer.de}}%
	\thanks{Marco F. Huber is with the Institute of Industrial Manufacturing and Management IFF, University of Stuttgart, 70569 Stuttgart, Germany {\tt\small marco.huber@ieee.org}}%
}

\begin{document}

	\maketitle
	\thispagestyle{empty}
	\pagestyle{empty}

	\begin{abstract}
		Identifying parameters in a system of nonlinear, ordinary differential equations is vital for designing a robust controller. However, if the system is stochastic in its nature or if only noisy measurements are available, standard optimization algorithms for system identification usually fail. We present a new approach that combines the recent advances in physics-informed neural networks and the well-known achievements of Kalman filters in order to find parameters in a continuous-time system with noisy measurements. In doing so, our approach allows estimating the parameters together with the mean value and covariance matrix of the system's state vector. We show that the method works for complex systems by identifying the parameters of a double pendulum.
	\end{abstract}

	\section{Introduction}
	\label{sec:introduction}
	Controlling a dynamical system in a safe manner requires a model that describes the system properties precisely. \Acp{ODE} are often used to satisfy this requirement. Besides setting up the corresponding equation operators, it is also inevitable to identify the real-valued coefficients that define the characteristics of the system. Estimating these parameters by using measurements is termed as ``inverse problem'' and can be a difficult task, depending on the system's complexity. This work presents a new method that is capable of identifying unknown parameters in a nonlinear \ac{ODE} system, based on noisy measurements by using an \ac{EKBF} in a machine learning framework.
	
	For linear systems, the subspace-based state space identification methods are well established. They aim at finding a linear state space model by using a regularized least-squares algorithm~\cite{Ljung.2010}. If the system comprises nonlinear behavior, the most straightforward solution approaches for parameter identification are standard minimization techniques like gradient-based~\cite{Mitra.2019} or gradient-free~\cite{Akman.2018} methods. For a system with noisy measurements, the problem becomes even more difficult and requires incorporating stochastic moments in the optimization. Raue et. al. summarize their experiences of fitting measurements of biological systems to their corresponding \Ac{ODE} system by maximizing a log-likelihood function that comprises a normally distributed measurement noise~\cite{Raue.2013}. 
	However, these methods require a numerical solution of the \ac{ODE}, repeatedly for each optimization iteration. Besides being very time consuming, this approach often fails because of the system's nonlinearity, the noise influence or an unstable behavior in the numerical solution~\cite{Liang.2008}.
	
	A possibility to circumvent the problem with a machine learning approach is described in \cite{Stubberud1995} by using a neural network to improve a Kalman filter system in order to obtain a better state estimate. Though, this does not give us the actual system parameter values but compensates for model errors. In 2017, Raissi et. al. presented how \acp{PINN} can be trained by using modern automatic differentiation frameworks~\cite{Raissi.28.11.2017}. The approach utilizes deep neural networks to discover and solve nonlinear differential equation systems. This is achieved by training a neural network to represent an approximate solution to the differential equation. The method also enables a parameter search, by including the unknown parameters as additional network weights. The concept has been applied in numerous research fields, e.g., mechanics \cite{Misyris.2020}, thermodynamics~\cite{Niaki.2021} or in chemical reaction equations~\cite{Ji.2021}. \acp{PINN} also enable the possibility to include stochastic behavior in the modeling process. Recently, this has been addressed by O'Leary et. al. who incorporate a mean value of the state and its covariance matrix in the framework, leveraging it to a \ac{SPINN}~\cite{OLeary.03.09.2021}. The authors do so by propagating the first two central moments of a state variable through the known differential equation systems. Afterwards a neural network is trained in order to match these estimated central moments to measured ones. 
	However, the authors do not address the problem of identifying parameters in the system. Another option is to use a \ac{BNN} in \iac{PINN} environment which allows an embedding of uncertainty and, hence, the usage of stochastic differential equations. Yang et. al. use a \ac{BNN}~\cite{Yang.2021} to include noisy data into a partial differential equation problem in order to solve as well as identify the system~\cite{Yang.2021}. However, \acp{BNN} are often not capable of achieving the same approximation accuracy as standard neural networks and are significantly more difficult to train.
	
	In this paper, we present a new physics-informed machine learning approach that we call \acfi{KBINN}. A \acl{KBF} incorporates two \acp{ODE} that describe the temporal evolution of the mean value and the covariance matrix of the system's state. In our method, we include two neural networks that are implemented in \iac{PINN} framework in order to approximate a solution to the Kalman-Bucy equations. This allows an implicit identification of unknown system parameters by incorporating them into the network training. The rest of the paper is organized as follows: In Section~\ref{sec:problem_formulation}, we give a short mathematical formulation of the problem. Section~\ref{sec:preliminaries} introduces the \acfi{EKBF} and gives a short summary of neural networks. Section~\ref{sec:kbinn} describes the \ac{KBINN} method, followed by validation experiments in Section~\ref{sec:experiments}. We discuss the strengths and limitations of our method in Section~\ref{sec:discussion} and close the paper with a conclusion in Section~\ref{sec:conclusions_and_outlook}.
	
	\section{Problem Formulation}
	\label{sec:problem_formulation}
	The state space representation of a continuous-time, nonlinear, dynamic and time-variant system of rank $n \in \mathbb{N}$ is defined by means of
	\begin{equation} \label{eq:ss_representation}
		\begin{split}
			\bm{\dot{x}}(t) &= \bm{f}\left( \bm{x}(t), \bm{u}(t), \bm{w}(t),t,\bm{\theta} \right) \\
			\bm{y}(t) &= \bm{g}\left( \bm{x}(t), \bm{u}(t), \bm{v}(t),t \right) ~,
		\end{split}
	\end{equation}
	where $f(\cdot)$ is the nonlinear ODE and $g(\cdot)$ is the measurement function. Both are assumed to be known, except for a set of unknown parameters. Furthermore, $\bm{x}(t) \in \mathbb{R}^n$ with $t\ge 0$ denotes the state vector, $\bm{u}(t) \in \mathbb{R}^p$ and $\bm{y}(t) \in \mathbb{R}^q$ denote the input and output signal with dimensions $p,q \in \mathbb{N}$, respectively. The vectors $\bm{w}(t) \in \mathbb{R}^n$ and $\bm{v}(t) \in \mathbb{R}^q$ denote white process noise and white measurement noise, respectively, which are both assumed to be zero-mean Gaussian with covariance matrices $\bm{Q}(t) \in \mathbb{R}^{n\times n}$ and $\bm{R}(t) \in \mathbb{R}^{q\times q}$, respectively. This induces the state $\bm{x}(t)$ to be a random variable as well. $\bm{\theta} \in \mathbb{R}^d$ denotes a vector of $d \in \mathbb{N}$ unknown parameters of the ODE system. If we acquire noisy measurements $\overline{\bm{y}}(t_i)$ of the system at $N$ discrete time steps $t_i$ with $i = 1,\ldots,N$, the aim of our method is to find a parameter vector $\bm{\theta}^*$ that minimizes the objective function
	\begin{equation}
		\label{eq:ss_obj_function}
		\bm{\theta}^* =  \arg\min_{\bm{\theta}} \left( \bm{y}(t_i) - \overline{\bm{y}}(t_i) \right)^2 .
	\end{equation}
	Standard minimization algorithms require to solve the ODE in Eq.~\eqref{eq:ss_representation} at each iteration numerically in order to minimize Eq.~\eqref{eq:ss_obj_function}. As has been mentioned in Section~\ref{sec:introduction}, these algorithms often fail due to the necessary small step width which is imposed by the system's nonlinearity or the noise.
	
	\section{Preliminaries}
	\label{sec:preliminaries}
	In this section, we briefly introduce the \ac{EKBF}, which is based on the well-known Kalman filter, applied to the nonlinear and continuous-time case. Afterwards, we give a short introduction to artificial neural networks.
	
	\subsection{Extended Kalman-Bucy Filter}
	It is obvious from Eq.~\eqref{eq:ss_representation} that the state $\bm{x}$ is not directly measurable. To compensate for this problem, Rudolf E. Kalman introduced in 1960 the Kalman filter first for discrete-time, linear systems and a year later for the continuous-time case, together with Richard S. Bucy \cite{Kalman.1961}. This concept allows estimating the mean value and the covariance matrix of the system's state. If the noise is assumed to be Gaussian, the first two central moments are sufficient to describe the state's probability distribution exactly. If a nonlinear system is considered, it is necessary to perform linearizations at each time step which leads to the \ac{EKBF}.  It is composed of two initial value problems that comprises the state's estimated mean value
	\begin{multline} \label{eq:kb_mean}
		\bm{\dot{\hat{x}}}(t) = \bm{f}\left(\bm{\hat{x}}(t),\bm{u}(t),\bm{0},t \right) \\
		+ \bm{K}(t) \cdot \left( \bm{\overline{y}}(t) - \bm{g}(\bm{\hat{x}}(t), \bm{u}(t),\bm{0},t) \right)
	\end{multline}
	with a known initial value $\bm{\hat{x}}(0) = \bm{\hat{x}}_0$ and a Kalman Gain
	\begin{equation} \label{eq:kb_gain}
		\bm{K}(t) = \hat{\bm{P}}(t) \cdot \bm{\hat{C}}^\mathrm{T}(t) \cdot \bm{\hat{R}}^{-1}(t)
	\end{equation}
	as well as its covariance matrix
	\begin{multline} \label{eq:kb_cov}
		\dot{\hat{\bm{P}}}(t) = \bm{\hat{A}}(t)\bm{\hat{P}}(t) + \bm{\hat{P}}(t)\bm{\hat{A}}(t)^\mathrm{T} \\ - \bm{\hat{P}}(t)\bm{\hat{C}}^\mathrm{T}(t)\bm{\hat{R}}^{-1}(t) \bm{\hat{C}}(t) \bm{\hat{P}}(t) + \bm{\hat{Q}}(t)
	\end{multline}
	with a known initial value $\bm{\hat{P}}(0) = \bm{\hat{P}}_0$. In Eq.~\eqref{eq:kb_gain} and \eqref{eq:kb_cov}, the involved matrices are obtained by linearization of Eq.~\eqref{eq:ss_representation} according to
	\begin{equation}
		\label{eq:linearizations}
		\begin{split}
			\bm{\hat{A}}(t) &= \left. \frac{\partial \bm{f}(\bm{x},\bm{u}, \bm{w}, t)}{\partial \bm{x}(t)} \right|_{\wedge},~
			\bm{\hat{C}}(t) = \left. \frac{\partial \bm{g}(\bm{x},\bm{u}, \bm{v}, t)}{\partial \bm{x}(t)} \right|_{\wedge}, \\
			\bm{\hat{G}}(t) &=  \left. \frac{\partial \bm{f}(\bm{x},\bm{u}, \bm{w}, t)}{\partial \bm{w}(t)} \right|_{\wedge},~
			\bm{\hat{V}}(t) = \left. \frac{\partial \bm{g}(\bm{x},\bm{u}, \bm{v}, t)}{\partial \bm{v}(t)} \right|_{\wedge}.
		\end{split}
	\end{equation}
	The $\wedge$-symbol denotes that the linearization is performed repeatedly for each new mean value $\bm{\hat{x}}(t)$. This also allows obtaining the noise covariance matrices
	\begin{equation}
		\begin{split}
			\bm{\hat{Q}}(t) &= \bm{\hat{G}}(t) \cdot \bm{Q}(t) \cdot \bm{\hat{G}}^\mathrm{T}(t)~, \\
			\bm{\hat{R}}(t) &= \bm{\hat{V}}(t) \cdot \bm{R}(t) \cdot \bm{\hat{V}}^\mathrm{T}(t) 
		\end{split}
	\end{equation}
	in Eq.~\eqref{eq:kb_gain} and \eqref{eq:kb_cov}, respectively. 
	Note that we omitted $\bm{\theta}$ from Eq.~\eqref{eq:ss_representation} for introducing the \ac{EKBF}, since the filtering problem does not aim at identifying parameters in a system, but only enables us to calculate the state's mean value and its covariance matrix in a stochastic environment.
	The necessity of performing a linearization for every new state usually leads to a considerable computing effort which lowers the attractiveness of the \ac{EKBF} for many applications. But the recent advances in automatic differentiation and its usage in neural networks make this method perfect for our purpose.
	
	\subsection{Neural Networks}
	A neural network is a type of machine learning algorithm that maps an input signal of rank $\iota$ to an output signal of rank $\kappa$ by approximating a desired function. In its simplest form, it consists of at least three layers: the input layer comprises $\iota$ neurons and does not perform any transformations but only distributes the input signal to the successive layer. The hidden layer comprises a variable count of neurons. Each performs a weighted, nonlinear transformation by means of
	\begin{equation} \label{eq:nn_transformation}
		o = \sigma \left( \sum_{k=1}^l i_k w_k + w_0 \right) = \sigma \left( \bm{i}^\mathrm{T} \cdot \bm w \right) ~.
	\end{equation}
	Here, $\bm{i} = \left[1,i_1,\ldots,i_l \right]^\mathrm{T} \in \mathbb{R}^{l+1}$ denotes the output of the previous layer with $l$ neurons and $\bm{w}= \left[w_0,w_1,\ldots,w_l \right]^\mathrm{T} \in \mathbb{R}^{l+1}$ denotes a weighting vector. If the neural network is used to approximate a nonlinear behavior, the activation function $\sigma \left( \cdot \right) \colon \mathbb{R} \to \mathbb{R}$ needs to be of nonlinear nature as well. A neural network can comprise several successive hidden layers. The last layer, called output layer, is composed of $\kappa$ neurons and also performs the calculation of Eq.~\eqref{eq:nn_transformation}. It can be shown that a neural network with at least one hidden layer but an arbitrary count of neurons and bounded, continuous, non-constant activation functions is able to approximate any given function~\cite{Hornik.1989}.
	
	\section{Physics-Informed Neural Networks for System Identification}
	\label{sec:kbinn}
	In this section, we first introduce the concept of \acp{PINN} in order to perform an identification of an \ac{ODE} system. The idea of using \acp{PINN} as a tool for system identification has first been mentioned in~\cite{Raissi.28.11.2017}. We extend this approach by incorporating noise signals and take advantage of an \ac{EKBF} in order to estimate a system's mean value and covariance matrix. This allows estimating parameters even if there are only noisy measurements available or if the model is imprecise.
	
	\subsection{Physics-Informed Neural Network}
	\label{sec:pinn}
	\acp{PINN} consider an initial value problem of the form
	\begin{equation} \label{eq:pinn_ss}
		\begin{split}
			\bm{\dot{x}}(t) &= \bm{f}\left( \bm{x}(t),\bm{u}(t), t, \bm{\theta} \right) \\
			\bm{y}(t) &= \bm{g}\left( \bm{x}(t),\bm{u}(t), t \right)
		\end{split}
	\end{equation}
	with equivalent variables to Eq.~\eqref{eq:ss_representation} and an initial value $\bm{x}(0) = \bm{x}_0$. Note that we do not consider any noise in this case. Let $\bm{x}(t)$ be an (approximate) solution to Eq.~\eqref{eq:pinn_ss}, depending on the existence. The idea is to train a neural network $\pi(t, \bm{W}) \colon \mathbb{R}^{p+1} \to \mathbb{R}^n$ that approximates the solution $\bm{x}(t)$ by adjusting the network weights which are summarized in a matrix $\bm{W}$. Parallel to the training, a vector $\hat{\bm{\theta}}\in \mathbb{R}^d$, that comprises an estimate of the original parameters $\bm{\theta}$, becomes optimized. We do so by setting up a loss function according~to
	\begin{multline}\label{eq:pinn_loss}
		J_\mathrm{PINN} = \sum_{i=1}^N\left( \frac{\mathrm{d}}{\mathrm{d}t} \pi(t_i) - \bm{f}\left( \pi(t_i),\bm{u}, t_i, \hat{\bm{\theta}} \right) \right)^2 \\ + \left( g\left(\pi(t_i), \bm{u},t_i \right) - \bm{\overline{y}}_i \right)^2 ~,
	\end{multline}
	with $\bm{\overline{y}}_i$ denoting a measurement of the system at time $t_i$. For the sake of readability, we abbreviated the full argument list of $\pi(t,\bm{W})$. Subsequently, we execute a minimization algorithm in order to find the optimal values for the weight matrix $\bm{W}$ and the estimated parameter vector $\hat{\bm{\theta}}$. If successful, minimizing the loss function in Eq.~\eqref{eq:pinn_loss} yields the true parameters $\bm{\theta}$ and a neural network that features the same derivative and initial value as $\bm{x}(t)$.
	
	\subsection{Kalman-Bucy-Informed Neural Network}
	\begin{figure*}[t!] 
		\centering
		\begin{tikzpicture}

\draw[draw=black] (0,0) rectangle ++(6,-1.5);
\node[draw,anchor=north west] at (0,0) {\textbf{Mean network}};
\node[draw,anchor=south](xi) at (2,-1.5) {$\xi(t,\bm{W}_x)$};
\node[draw,anchor=south](xi_dot) at (4.5,-1.5) {$\frac{\mathrm{d}}{\mathrm{d}t}\xi(t,\bm{W}_x)$};

\node[anchor=south](x) at (2,-2.5) {$\hat{\bm{x}}(t)$};
\node[anchor=south](x_dot) at (4.5,-2.5) {$\dot{\hat{\bm{x}}}(t)$};

\draw[->] (xi) -- (x);
\draw[->] (xi_dot) -- (x_dot);

\node[draw, anchor=north](g) at (0.5,-2.95) {$\bm{g}(\hat{\bm{x}}(t),0,v)$};

\coordinate(en1) at (2,-2.75);
\coordinate(en2) at (0.5,-2.75);
\draw[->] (x) -- (en1) -- (en2) -- (g);

\draw[draw=black] (7,0) rectangle ++(6,-1.5);
\node[draw,anchor=north west] at (7,0) {\textbf{Covariance network}};
\node[draw,anchor=south](psi) at (8.5,-1.5) {$\psi(t,\bm{W}_P)$};
\node[draw,anchor=south](psi_dot) at (11.5,-1.5) {$\frac{\mathrm{d}}{\mathrm{d}t}\psi(t,\bm{W}_P)$};

\node[anchor=south](P) at (8.5,-2.5) {$\hat{\bm{P}}(t)$};
\node[anchor=south](P_dot) at (11.5,-2.5) {$\dot{\hat{\bm{P}}}(t)$};

\draw[->] (psi) -- (P);
\draw[->] (psi_dot) -- (P_dot);

\draw[draw=black](0,-3.75) rectangle++(13,-0.5);
\node[anchor=center] at (6.5,-4) {Loss $J$};

\draw[->](x) -- (2,-3.75);
\draw[->](x_dot) -- (4.5,-3.75);
\draw[->](g) -- (0.5,-3.75);

\draw[->](P) -- (8.5,-3.75);
\draw[->](P_dot) -- (11.5,-3.75);

\node[anchor=center](time) at (6.5,0.25) {Time $t$};
\draw[->] (time) -- (3,0.25) -- (3,0);
\draw[->] (time) -- (10,0.25) -- (10,0);

\draw[-stealth, dashed] (0,-4) -- (-1,-4) -- (-1,-0.75) -- (0,-0.75);
\node[anchor=west] at (-1,-2.25) {$\Delta \bm{W}_x$};

\draw[-stealth,dashed] (13,-4) -- (14,-4) -- (14,-0.75) -- (13,-0.75);
\node[anchor=east] at (14,-2.25) {$\Delta \bm{W}_P$};

\node[draw](theta) at (6.5,-4.75){$\bm{\theta}$};
\node() at (4.25,-4.5){$\hat{\bm{\theta}}$};
\node(theta_delta) at (8.75,-4.5){$\Delta\bm{\theta}$};

\draw[-stealth, dashed] (11,-4.25) -- (11,-4.75) -- (theta);
\draw[->] (theta) -- (2,-4.75) -- (2,-4.25);
\end{tikzpicture}
		\caption{Sketch of the KBINN. Two neural networks are trained in order to approximate \iac{PINN}-based solution for the \ac{EKBF}. The output of both networks as well as the estimated parameters $\hat{\bm{\theta}}$ are fed into the loss function $J$. Changes of the training parameters are indicated by dashed lines and a $\Delta$ in front of the variable.}
		\label{fig:approach}
	\end{figure*}
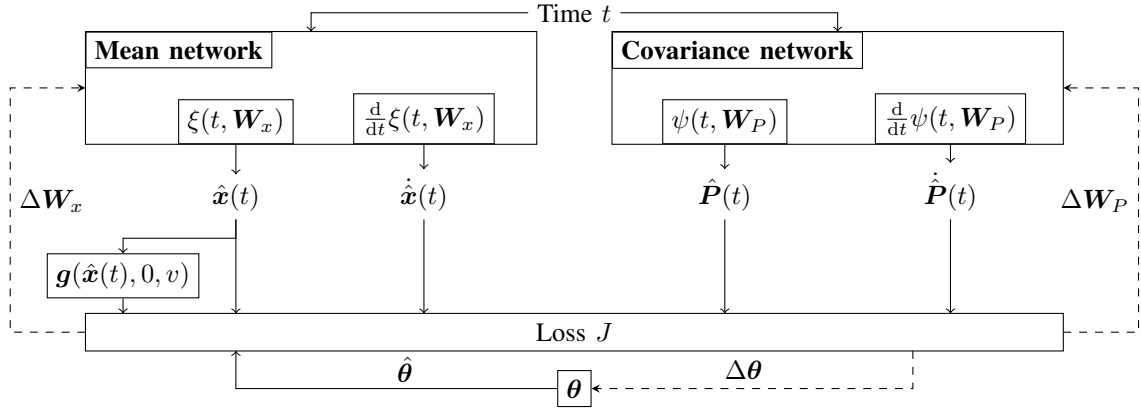
	In our approach, we consider a system with noise as in Eq.~\eqref{eq:ss_representation}. They key idea is to implement \iac{PINN} that approximates the solution of the initial value problems from Eq.~\eqref{eq:kb_mean} and Eq.~\eqref{eq:kb_cov}. This results in a method that allows estimating the parameters of a stochastic \ac{ODE} system given noisy measurements and a possibly erroneous model. We do so by constructing two neural networks that receive a time signal as input and predict the state's mean value $\bm{\hat{x}}(t)$ as well as its covariance matrix $\bm{\hat{P}}(t)$, respectively. Fig.~\ref{fig:approach} depicts a sketch of our approach. The \emph{mean network} $\xi(t, \bm{W}_x)$  with weight matrix $\bm{W}_x$ approximates the state's mean value $\hat{\bm{x}}(t)$. Its counter part for the covariance matrix is the \emph{covariance network} $\psi(t,\bm{W}_P)$ with a weighting matrix $\bm{W}_P$. The vector $\hat{\bm{\theta}}$ contains the estimated parameters. We optimize both weighting matrices $\bm{W}_x$ and $\bm{W}_P$, together with $\hat{\bm{\theta}}$ simultaneously by minimizing a loss function $J$ that is composed of three weighted terms according to
	\begin{equation} \label{eq:main_loss}
		J = \sum_{i=1}^N \left( \alpha_1 L_{1,i} + \alpha_2 L_{2,i} + \alpha_3 L_{3,i}\right)
	\end{equation}
	with weights $\alpha_j > 0$, $j \in \left\{1,2,3 \right\}$ that can be chosen arbitrarily. To improve readability, we again drop the weighting matrices from both neural network functions. The first term $L_{1,i}$ in \eqref{eq:main_loss} is given by
	\begin{multline}
		L_{1,i} = \lnorm\xi(0) - \bm{x}_0 \rnorm_2 + \lnorm \dot{\xi}(t_i) - \bm{\Xi}(\xi,t_i) \rnorm_2
	\end{multline}
	with
	\begin{multline}
		\bm{\Xi}(\xi,t_i) = \bm{f}\left( \xi(t_i),\bm{u},\bm{0}, t_i, \hat{\bm{\theta}} \right) \\
		+ \bm{K}(t_i) \cdot \left(\overline{\bm{y}}_i - \hat{\bm{y}}(t_i) \right) .
	\end{multline}
	and $\bm{K}(t_i) = \psi(t_i) \cdot \bm{\hat{C}}^\mathrm{T}(t_i) \cdot \bm{\hat{R}}^{-1}(t_i)$ as a Kalman gain. $\hat{\bm{y}}(t_i) = \bm{g}(\xi(t_i),\bm{u},\bm{0},t_i)$ describes the estimated output and $\lnorm \cdot \rnorm_2$ denotes the Euclidean norm. Hence, $L_{1,i}$ utilizes  Eq.~\eqref{eq:kb_mean} and \eqref{eq:kb_gain} by replacing $\hat{\bm{x}}(t)$ and $\hat{\bm{P}}(t)$ with the corresponding neural networks. It converges to $0$ if $\dot{\xi}(t)$ and $\xi(0)$ approach $\bm{\dot{\hat{x}}}(t)$ and $\bm{x}_0$. The second term $L_{2,i}$ is an equivalent to $L_{1,i}$, applied to the covariance matrix. It is defined by means of
	\begin{multline}
		L_{2,i} = \lnorm \psi(0) - \hat{\bm{P}}_0 \rnorm_\mathrm{F} + \lnorm \dot{\psi}(t_i) - \bm{\Psi}(\psi,t_i) \rnorm_\mathrm{F}
	\end{multline}
	with the Frobenius norm $\lnorm \cdot \rnorm_\mathrm{F}$ and
	\begin{multline}
		\bm{\Psi}(\psi,t_i) = \bm{\hat{A}}(t_i)\psi(t_i) + \psi(t_i)\bm{\hat{A}}(t_i)^\mathrm{T}  \\ - \psi(t_i)\bm{\hat{C}}^\mathrm{T}(t_i)\bm{\hat{R}}^{-1}(t_i) \bm{\hat{C}}(t_i) \psi(t_i) + \bm{\hat{Q}}(t_i)~,
	\end{multline}
	which utilizes Eq.~\eqref{eq:kb_cov} where we substitute the covariance matrix $\hat{\bm{P}}$ by the corresponding neural network $\psi(t_i)$.
	The third term $L_{3,i}$ aims at keeping the original measurement values close to the uncertainty interval that is spanned by the mean and its covariance by incorporating a maximum likelihood estimator. Since we assume the measurement noise to be additive and Gaussian, we use a normal distribution by means of
	\begin{equation}
		p_j\left(\overline{\bm{y}}_{j}(t_i); \xi(t_i),\psi(t_i)\right) = \mathcal{N}\left(\overline{\bm{y}}_{j}(t_i); \bm{\mu}_j(t_i), \bm{\sigma}^{2}_j(t_i)\right)
	\end{equation}
	with $j = 1,\ldots,q$ for each output dimension. We calculate the mean value by means of $\bm{\mu}(t_i) = \mathrm{mean}\left(\bm{g}\left(\bm{x}, \bm{u}, \bm{v}, t_i\right)\right)$ and the variance according to $\bm{\sigma}^{2}(t_i) = \mathrm{var}\left(\bm{g}\left(\bm{x}, \bm{u}, \bm{v}, t_i \right)\right)$ with $\bm{x}(t_i) \sim \mathcal{N}\left(\bm{x}(t_i); \xi(t_i), \psi(t_i) \right)$. Both, the mean and the variance can be calculated in closed form if the measurement function exhibits a special type, e.g., linear, polynomial or trigonometric \cite{Huber.2015}. In other cases, approximations are necessary, e.g. sampling or numerical integration. We then obtain the loss $L_{3,i}$ by using the negative natural logarithm
	\begin{equation}
		L_{3,i} = -\sum_{j=1}^q \log \left( p_j\left(\overline{\bm{y}}_{j}(t_i);\xi(t_i),\psi(t_i)\right)  \right) ~.
	\end{equation}
	
	\subsection{Implementation}
	\label{sec:implementation}
	In this section, we explain the implementation details  of our approach. For both neural networks, all activation functions need to be at least twice continuous differentiable, because we use the network's temporal derivative and a gradient-based weight optimization. The neural network $\xi(t,\bm{W}_x)$ possesses one input neuron and $n$ output neurons, with $n$ denoting the system's rank. The neuron and layer count in the hidden layers depends on the particular use case. We choose the hyperbolic tangent as activation function for all hidden neurons. The output layer comprises a linear activation function. The covariance network $\psi(t,\bm{W}_P)$ is slightly more complex. While the architecture of input and hidden layers are identical to $\xi(t,\bm{W}_x)$, we must ensure that the network's output comprises a symmetric positive semi-definite matrix of rank $n$. We do so by constructing $(n+1) \cdot \nicefrac{n}{2}$ output neurons and build an upper triangular matrix. Afterwards, we multiply the upper triangular matrix with its transposed version.
	
	After setting up the neural networks as well as $\hat{\bm{\theta}}$, the weights $\bm{W}_x$ and $\bm{W}_P$ as well as the parameters are optimized simultaneously by minimizing the loss function of Eq.~\eqref{eq:main_loss}. Therefore we use the backpropagation algorithm \cite{Rumelhart.1986}, which utilizes a gradient-based optimization to efficiently fit the weights of a neural network.
	
	\section{Experiment: Identification of a Double Pendulum System}
	\label{sec:experiments}
	We validate the proposed \ac{KBINN} by solving the inverse problem of a double pendulum system. For comparison, we employ \MATLAB{}'s system identification toolbox, which is dedicated to identify parameters by using system measurements. The function \texttt{nlgreyest} allows estimating model parameters of a nonlinear \ac{ODE} by iteratively solving it numerically and using a gradient-based optimization algorithm. 
	
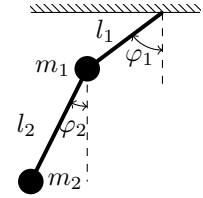
\begin{figure}[t!]
		\centering
		\begin{tikzpicture}
\coordinate(anchor) at (1.5,2.25);
\coordinate(en1) at (1.5,1.25);
\coordinate(m1) at (0.5,1.5);
\coordinate(en2) at (0.5,0);
\coordinate(m2) at (-0.25,0);

\draw[draw=none, pattern=north west lines, pattern color=black] (-0.25,2.25) rectangle ++(2.25,0.1);
\draw[] (-0.25,2.25) -- (2,2.25);

\node[circle,inner sep=1.2mm,fill=black] (a) at (m1) {};
\node[circle,inner sep=1.2mm,fill=black] (b) at (m2) {};

\draw[line width=1.5pt] (m1)--(m2);
\draw[,line width=1.5pt] (m1)--(anchor);
\draw[dashed] (anchor) -- (en1);
\draw[dashed] (m1) -- (en2);

\pic [draw=black, <->, "$\varphi_1$", angle eccentricity=1.35] {angle = m1--anchor--en1};
\pic [draw=black, <->, "$\varphi_2$", angle eccentricity=1.65] {angle = m2--m1--en2};

\node[anchor=east] () at (0.4,1.5) {$m_1$};
\node[anchor=west] () at (-0.15,0) {$m_2$};
\node[anchor=east] () at (-0.05,0.75) {$l_2$};
\node[anchor=east] () at (1.0,2.0) {$l_1$};
\end{tikzpicture}	
		\caption{Sketch of a double pendulum.}
		\label{fig:dp_sketch}
	\end{figure}

	A double pendulum is composed of a conventional pendulum with mass $m_1$ and rod length $l_1$, extended by another one with values $m_2$ and $l_2$. Generally, the system's parameters are hard to identify because of the chaotic motion behavior that lets small deviations in the initial position cause substantial trajectory differences. Fig.~\ref{fig:dp_sketch} depicts a sketch of the system. 
	The state is defined by the pendulum angles and their angular velocities $\bm{x} = \begin{bmatrix} \varphi_1 & \dot{\varphi_1} & \varphi_2 & \dot{\varphi_2} \end{bmatrix}^\mathrm{T}$. We define $c_{12} = \cos(\varphi_1 - \varphi_2)$, $s_{12} = \sin(\varphi_1 - \varphi_2)$ and $M = \frac{m_2}{m_1 + m_2}$. The corresponding ODE is given by~\cite{Stachowiak.2006}
	\begin{equation} \label{eq:double_pendulum}
		\begin{split}
			\ddot{\varphi}_1 &= -M \frac{l_2}{l_1} (\ddot{\varphi}_2 c_{12} + \dot{\varphi}_2^2 s_{12}) - \frac{g}{l_1} \sin(\varphi_1) \\
			\ddot{\varphi}_2 &= - \frac{l_1}{l_2} (\ddot{\varphi}_1 c_{12} - \dot{\varphi}_1^2 s_{12}) - \frac{g}{l_2} \sin(\varphi_2)
		\end{split}
	\end{equation}
	with the gravitational constant $g=\SI{9.81}{m/s^2}$. All state variables are time dependent. Our objective is to find the parameter vector $\bm{\theta} = \begin{bmatrix}l_1 & l_2 & M \end{bmatrix}$ that comprises the pendulum's lengths and masses. Since the KBINN is capable of dealing with small modeling errors, we exacerbate the problem by adding a damping term to Eq.~\ref{eq:double_pendulum} which yields
	\begin{equation} \label{eq:double_pendulum_damped}
		\begin{split}
			\ddot{\varphi}_1 &= -M \frac{l_2}{l_1} (\ddot{\varphi}_2 c_{12} + \dot{\varphi}_2^2 s_{12}) - \frac{g}{l_1} \sin(\varphi_1) - 0.05\dot{\varphi_1} \\
			\ddot{\varphi}_2 &= - \frac{l_1}{l_2} (\ddot{\varphi}_1 c_{12} - \dot{\varphi}_1^2 s_{12}) - \frac{g}{l_2} \sin(\varphi_2) - 0.05\dot{\varphi_2}~.
		\end{split}
	\end{equation}
	We assume all state variables in $\bm{x}$ to be measurable after adding a noise according to $\bm{y}(t) = \bm{x}(t) + \bm{v}(t)$ with $\bm{v}(t)$ being zero-mean Gaussian that comprises a covariance matrix of $\bm{R}(t) = \bm{I}\cdot 0.25$ with the identity matrix $\bm I$ of rank $n=4$.  Without assuming all states to be measurable, the parameter identification turns out to be too error-prone. Nevertheless, we successfully tested the algorithm on less complex systems (e.g.,
	an elastic pendulum) without a fully observable state vector. We use Eq.~\eqref{eq:double_pendulum_damped} to simulate system trajectories by solving the \ac{ODE} numerically and add noise as mentioned above to acquire $\overline{\bm{y}}(t_i)$ with $N = 3,000$ in a time interval of $3\,\mathrm{s}$ which yields a sampling frequency of~$\SI{1}{kHz}$.
	
\begin{figure*}[t!]
		\centering
		\begin{subfigure}[t]{0.49\linewidth}
			\raisebox{0.48cm}{\includegraphics[]{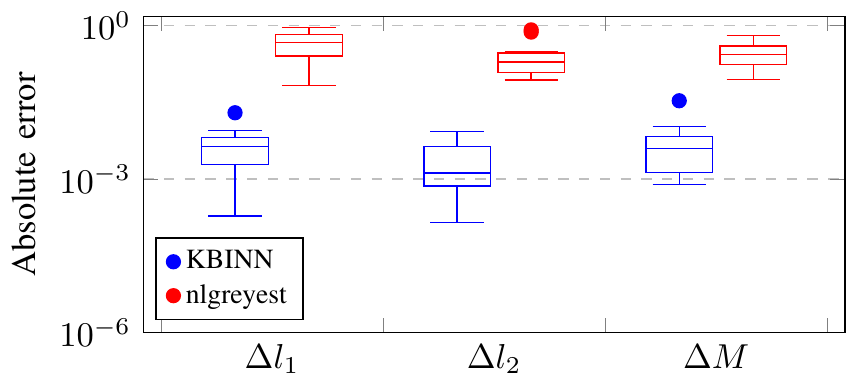}}
			\caption{Box plots of ten identification runs. The y-axis shows the absolute error between the original and identified parameter in accordance to its unit.}
		\end{subfigure}
		\begin{subfigure}[t]{0.49\linewidth}
			\includegraphics[]{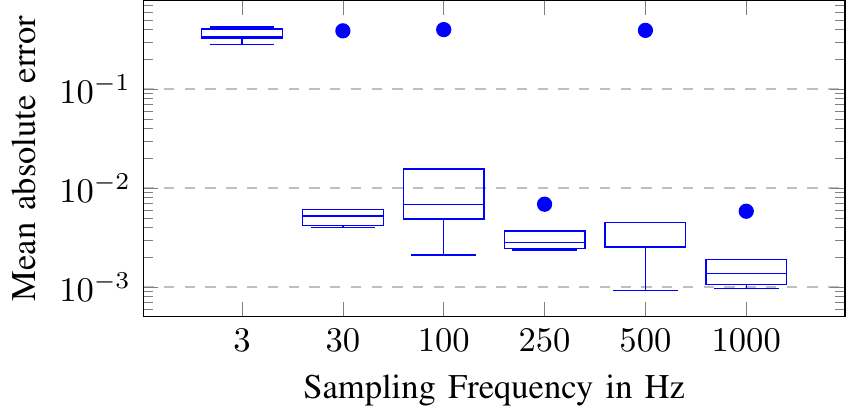}
			\caption{Box plots of five identification runs per sampling frequency. Each data point describes the average error over all parameters $\Delta l_1$, $\Delta l_2$, $\Delta M$.}
		\end{subfigure}
		\caption{Box plots of several identification runs to highlight the performance regarding (a) the estimate error for each parameter and (b) the performance as a function of the sampling frequency.}
		\label{fig:box_plots}
	\end{figure*}

	\begin{table}[t]
		\centering
		\caption{Mean absolute errors and standard deviations of the identified parameters over ten identification runs.}
		\begin{tabular}{l| c c c}
			Method & $\Delta l_1$ & $\Delta l_2$ & $\Delta M$ \\
			\hline
			KBINN & $0.03 \pm 0.11$ & $0.02 \pm 0.05$ & $0.08 \pm 0.24$\\
			nlgreyest & $0.49 \pm 0.28$ & $0.30 \pm 0.25$ & $0.30 \pm 0.17$ \\
		\end{tabular}
		\label{tab:mean_absolute_identification_errors}
	\end{table}
	To prove the functionality of our algorithm statistically, we perform ten system identification runs by sampling a new parameter vector $\bm{\theta}$ from a uniform distribution $\unif_{[0,1]}$ for each trial. Additionally, we generate new initial state vectors $\bm{x}_0 = \begin{bmatrix} \varphi_{1_{0}} & \dot{\varphi}_{1_{0}} &\varphi_{2_{0}} &\dot{\varphi}_{2_{0}}  \end{bmatrix}$ by sampling $\varphi_{1_{0}},\varphi_{2_{0}}$ from another uniform distribution $\unif_{[-\pi/4,\pi/4]}$. The initial angular velocities are kept to $0$. We also use new noise samples and new initial weight matrices for each cycle. We measure the success of the method by calculating the absolute error between the identified parameters. Table~\ref{tab:mean_absolute_identification_errors} shows the mean absolute errors of all identification runs for each parameter $\Delta l_1$, $\Delta l_2$ and $\Delta M$. We can see that on average, the KBINN leads to a more precise estimate, compared to the \texttt{nlgreyest}-method. Additionally, Fig.~\ref{fig:box_plots}(a) shows a box plot of the absolute errors. The ten identification runs depict one outlier sample which is clearly visible in the boxplot. The remaining nine trials all lead to an estimation error that is close to $0$. The \texttt{nlgreyest}-function on the other hand leads to estimates that differ heavily from the original values.
	
	We observed that the KBINN-based identification method requires obtaining measurements at a high sampling frequency. This is caused by the linearizations of the \ac{EKBF} in Eq.~\eqref{eq:linearizations}. A higher sampling frequency comes along with more data which requires more computing power and time. Thus, we want to find out the minimal sampling frequency that still allows a good parameter identification. We obtain five parameter and initial state scenarios by sampling from the same uniform distributions, as we used before. Afterwards, we identify the parameters based on different sampling frequencies which start at $\SI{3}{Hz}$ and increase up to $\SI{1}{kHz}$. We then calculate the mean value of $\Delta l_1$, $\Delta l_2$ and $\Delta M$ to acquire the box plot shown in Fig.~\ref{fig:box_plots}(b). As expected, a higher sampling frequency leads to a smaller error, on average. For the double pendulum, at least $\SI{100}{Hz}$ are required to get a steady deviation of less than one percent.
	
\begin{figure*}[t]
		\centering
		\begin{subfigure}[b]{0.49\linewidth}
			\includegraphics[]{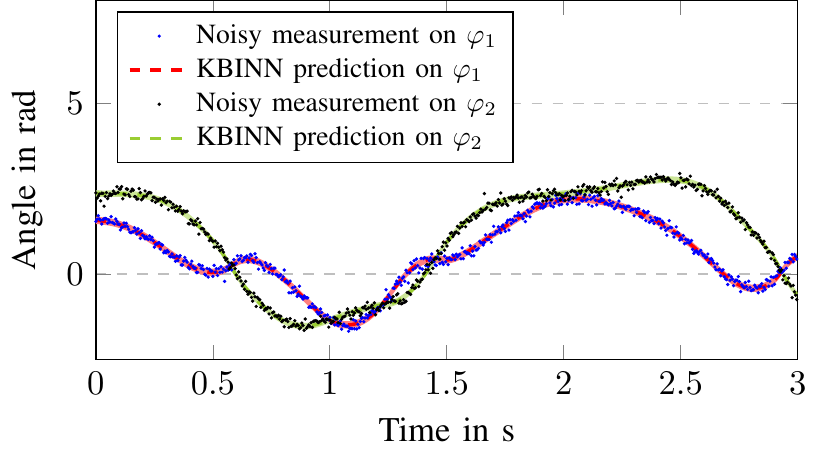}
			\caption{Trajectories and predictions of $\varphi_1$  and $\varphi_2$.}
		\end{subfigure}
		~
		\begin{subfigure}[b]{0.49\linewidth}
		\includegraphics[]{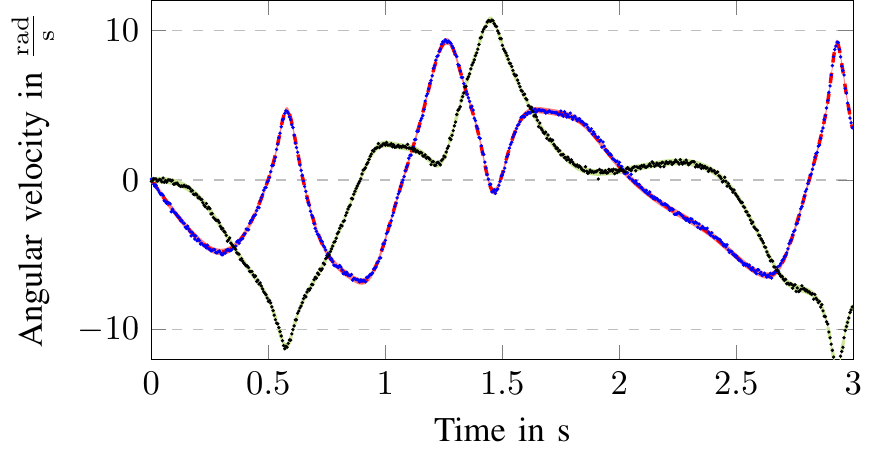}
		\caption{Trajectories and predictions of $\dot{\varphi}_1$ and $\dot{\varphi}_2$ with equivalent legend.}
		\end{subfigure}
		\vspace{-1mm}
		\caption{Plots of the simulated, noisy trajectories and the predictions of both neural networks $\xi(t)$ and $\psi(t)$ for the double pendulum. The uncertainty is depicted as colored $2\sigma$-interval, but it is barely visible as the values are close to zero.}
		\label{fig:dp_mean_cov_measurement}
	\end{figure*}

	In the following, we take one parameter scenario with $\bm{\theta} = \begin{bmatrix} \SI{0.6}{m} & \SI{0.9}{m} & 0.57 \end{bmatrix}$ as a showcase on how the KBINN networks behave. The parameter $M=0.57$ has been acquired by the masses $m_1 = \SI{0.3}{kg}$ and $m_2=\SI{0.4}{kg}$. Fig.~\ref{fig:dp_mean_cov_measurement} shows the noisy trajectories of all states in black. We also plot the predictions of the \ac{KBINN}'s mean and covariance networks $\xi(t)$ and $\psi(t)$. We can observe that the noisy trajectories stay mostly within the area that is spanned by the \ac{KBINN}'s predicted mean and variance.

	
	\section{Discussion}
	\label{sec:discussion}
	In this section, we discuss the advantages and limitations of our approach. As we showed in Section~\ref{sec:experiments}, we were always able to solve an inverse problem, although only noisy measurement data is given and the used ODE model contains modeling errors. In this paper, we only showed the identification results of three parameters, but further experiments indicate that our method scales really well with an increasing parameter count (\acp{ODE} with up to eight parameters were examined). We also want to stress that the identification is continuous-time as a whole and, thus, also works for systems that become unstable if discretized.
	
	However, the \ac{KBINN} method also has some limitations. The most severe one is the architecture choice of the neural networks.  The only possibility to find a good architecture is by trial and error or by using methods of neural architecture search \cite{white2021bananas}. Both, too few and too many neurons will lead to an unsuccessful identification. Furthermore, we sometimes observe a bad fit for the \ac{KBINN} networks in the last measurement points. The reason is a bad derivative estimate in the last time step that cannot influence the further trend anymore and, thus, does not affect the loss function heavily. We successfully experimented to solve this by extrapolating the measurements with a constant slope. Fortunately, the \ac{KBINN} indicates this false prediction by an increased variance. As we mentioned in Section~\ref{sec:experiments}, we also sometimes observe an outlier after several successful identification runs. We suspect that this occurs when the backpropagation algorithm is stuck in a local minimum during training. A possible work-around is to restart the training with a bigger step width. In our cases, we could correct the outliers by simply restarting the training with new initial weight matrices $\bm{W}_x$ and $\bm{W}_P$. Note that we did not perform this correction in our analysis in Section~\ref{sec:experiments}.
	
	\section{Conclusions and Outlook}
	\label{sec:conclusions_and_outlook}
	We introduced a new method to identify unknown parameters in an \ac{ODE} system with noisy measurements and imprecise models. The method uses recent advances in physics-informed machine learning by incorporating an \acl{EKBF}. We prove the performance of our method on a double pendulum system and compare the results to the corresponding \MATLAB-function. Future work is devoted to not only estimate the \ac{ODE}'s parameters but the entire structure of the equations.
	
	\section{Acknowledgements}
	This work was partially supported by the Baden-Wuerttemberg Ministry for Economic Affairs, Labour and Tourism (Project KI-Fortschrittszentrum
	\enquote{Lernende Syteme und Kognitive Robotik}) and the German Federal Ministry of Education and Research (Project Degrad-EL\textsuperscript{3}-Q).
	
	\balance
	%
	%
	\bibliographystyle{plain} 
	\bibliography{bibliography} 
	
\end{document}